\documentclass[submission,copyright,creativecommons]{eptcs}
\providecommand{\event}{VPT 2021} 
\usepackage{breakurl}             
\usepackage{underscore}           
\usepackage{textcomp}
\usepackage{makecell} 
\usepackage{graphicx}
\usepackage[final]{microtype}
\title{Conjectures, Tests and Proofs: \\ An Overview of Theory Exploration}
\author{Moa Johansson and Nicholas Smallbone
\institute{Chalmers University of Technology, Gothenburg, Sweden}
\email{moa.johansson@chalmers.se,  nicsma@chalmers.se}
}
\def\titlerunning{Conjectures, Tests and Proofs: An Overview of Theory Exploration}
\def\authorrunning{M. Johansson and N. Smallbone}
\begin{document}
\maketitle
\raggedbottom

\begin{abstract}
A key component of mathematical reasoning is the ability to formulate interesting conjectures about a problem domain at hand.
In this paper, we give a brief overview of a theory exploration system called QuickSpec, which is able to automatically discover interesting conjectures about a given set of functions. 
QuickSpec works by interleaving term generation with random testing to form candidate conjectures. This is made tractable by starting from small sizes and ensuring that only terms that are irreducible with respect to already discovered conjectures are considered. QuickSpec has been successfully applied to generate lemmas for automated inductive theorem proving as well as to generate specifications of functional programs.
We give an overview of typical use-cases of QuickSpec, as well as demonstrating how to easily connect it to a theorem prover of the user's choice.

\end{abstract}

\section{Introduction}
What makes a mathematical conjecture interesting and worth trying to prove as a lemma? For a human mathematician the motivation might be that the statement would make another proof shorter, clearer and easier to understand, or that the lemma captures, for instance, some common algebraic property of a structure of interest. Coming up with the right lemma in the right situation is sometimes described as a \emph{eureka step}: a sudden insight that makes a solution almost obvious. 

What about conjectures in programming? For a programmer, an interesting conjecture about a program at hand should shed light on its intended use and form part of a specification, which in turn can be used to for example generate test cases for the program, or even prove its correctness. Writing detailed formal specifications is however in practice something few programmers have the time or expertise to do.  

Automating these kind of creative steps in automated reasoning and program analysis is difficult both for symbolic and data-driven methods: Interesting lemmas might not fall out from simply applying deductive rules to axioms in a theorem prover; and when considering a new mathematical theory or program, we might not have a lot of previous knowledge in the domain, which makes it difficult to directly apply traditional machine learning techniques. 


In the context of automation of inductive proofs, there has been work on patching failed proof attempts by speculating lemmas using e.g. various \emph{proof critics}, ranging from simply generalising the goal by replacing common subterms with new variables, to more complex methods requiring specialised heuristics and search \cite{boyerMoore, productiveuse, isaplanner2}.  
In this article we will describe a different approach, \emph{theory exploration}, which instead of trying to construct lemmas from specific proof attempts, proceeds bottom-up: given a mathematical theory or program, what interesting conjectures can we come up with? 

We will demonstrate various use-cases of theory exploration centred 
around our tool QuickSpec \cite{quickspec}. QuickSpec generates conjectures about a set of functions and datatypes given by the user. It works by enumerating terms starting from small sizes up to a user specified limit, and then evaluating them at random values to determine if any terms appear to be equal.
To avoid uninteresting conjectures, and to manage the search space, only terms that have a unique normal form with respect to what has been discovered so far are kept.
To give the reader a flavour of what QuickSpec can do, 
Figure~\ref{fig:arith} shows the \emph{complete} output of QuickSpec on the theory of natural numbers with addition, multiplication and greatest common divisor, implemented as functions in Haskell. QuickSpec takes about 5 seconds to run on this example. Note that QuickSpec has no built-in knowledge of the laws of arithmetic, so everything in Figure~\ref{fig:arith} is invented from scratch.
\begin{figure}[htbp]
\begin{center}
\begin{verbatim}
 1. x+y = y+x               13. gcd(x,0) = x 
 2. x+0 = x                 14. gcd(x,1) = 1
 3. (x+y)+z=x+(y+z)         15. gcd(x,x*y) = x
 4. x*y = y*x               16. gcd(x,x+y) = gcd(x,y)
 5. x*0 = 0                 17. gcd(gcd(x,y),z) = gcd(x,gcd(y,z))
 6. x*1 = x                 18. gcd(x*y,x*z) = x*gcd(y,z)
 7. (x*y)*z = x*(y*z)       19. gcd(x*x,y*y) = gcd(x,y)*gcd(x,y)
 8. x*(y+y) = y*(x+x)       20. gcd(x*y,z+y) = gcd(x*z,z+y)
 9. x*(y+1) = x+(x*y)       21. gcd(x+x,y+y) = gcd(x,y)+gcd(x,y)
10. (x*y)+(x*z) = x*(y+z)   22. gcd(x+y,y+y) = gcd(x+x,x+y)
11. gcd(x,y) = gcd(y,x)     23. gcd(x*x,1+1) = gcd(x,1+1)
12. gcd(x,x) = x            
\end{verbatim}
\caption{Complete QuickSpec output on a small theory about arithmetic.}
\label{fig:arith}
\end{center}
\end{figure}
These conjectures include common properties such as commutativity, associativity, distributivity and identity properties, as well as more specific facts about \verb|gcd|, such as property 18 (about numbers having common factors), property 19 (about squares), the curious interchange property 20, and property 23 (which holds because 2 is prime).
Notice that after discovering for instance $x * 0 = 0$ (property 5), QuickSpec will not generate any terms where $x * 0$ is a subterm, as such a term would be reducible by applying property 5 as a rewrite rule, and therefore is not considered interesting. This is one of the key observations that makes term generation tractable (see \cite{quickspec} for full technical details and optimisations of the term generation algorithm). 
Note that QuickSpec itself does not prove any properties, but tests them thoroughly on randomly generated values, using Haskell's QuickCheck tool \cite{quickcheck}. Depending on the theory we are exploring we can of course pass the properties discovered by QuickSpec on to a suitable automated or interactive theorem prover. 

\section{A Small QuickSpec Tutorial}
In this section we demonstrate QuickSpec on two examples taken from programming: a data structure implementing maps from keys to values, and a library for describing musical compositions. The section serves as a tutorial for the aspiring QuickSpec user: we describe in detail how we ran QuickSpec, how and why we fine-tuned its settings for each example, and what obstacles we encountered along the way.

QuickSpec can be downloaded from \url{https://github.com/nick8325/quickspec}. The examples in this section are found in the following files in that repository:
\begin{itemize}
    \item Greatest common divisor: \verb|examples/GCD.hs|.
    \item Maps: \verb|examples/Map.hs|.
    \item Music: \verb|examples/Music/MusicQS.hs|.
\end{itemize}

\subsection{Greatest Common Divisor}
We first show how the conjectures in Figure~\ref{fig:arith} were produced. To run QuickSpec, the user first defines a \emph{signature}, which describes the set of functions we would like to explore (along with other configuration settings that we shall see later). The signature that we used for the arithmetic theory is shown in Figure~\ref{fig:arith-sig}.

\begin{figure}[h]
\begin{verbatim}
exampleSig = series [arithSig, gcdSig]
  where
    arithSig =
      [con "0" (0 :: Natural),
       con "1" (1 :: Natural),
       con "+" ((+) :: Natural -> Natural -> Natural),
       con "*" ((*) :: Natural -> Natural -> Natural)]
    gcdSig =
      [con "gcd" (gcd :: Natural -> Natural -> Natural)]
\end{verbatim}
\caption{A QuickSpec signature for the greatest common divisor function.}
\label{fig:arith-sig}
\end{figure}

The function \verb|con| is used in the signature to declare a Haskell constant or function. It takes two arguments: the first is the name of the constant, and the second is the constant itself. QuickSpec has no built-in constants that it automatically considers, so if we want to see conjectures involving \verb|0| and \verb|1|, we need to explicitly declare them. Type signatures are used to specialise our arithmetic functions to the natural numbers---any typeclass-polymorphic function must be specialised to a concrete type.

We also use the function \verb|series|. The expression \verb|series [arithSig, gcdSig]| tells QuickSpec that we would like to split the exploration into two parts: First we would like to explore the operations \verb|0|, \verb|1|, \verb|+| and \verb|*|, and once that is done we would like to add \verb|gcd| to the mix. By doing this, we see conjectures about \verb|gcd| only once all conjectures for the other operations have been generated.

Having written this signature, we can run QuickSpec by typing in \verb|quickSpec exampleSig| at the Haskell prompt, and the conjectures in Figure~\ref{fig:arith} are produced.

\subsection{Maps from Keys to Values}

The Haskell standard library defines a module \verb|Data.Map| which implements a map from keys to values. Figure \ref{fig:maps-types} shows a small part of the map API, which we will now explore with QuickSpec.

The type \verb|Map k v| represents a map from keys of type \verb|k| to values of type \verb|v|. The constant \verb|empty| is an empty map; \verb|insert| adds a key-value pair to a map; \verb|delete| removes a key-value pair given a key; and \verb|lookup| looks up a key to obtain a value, returning \verb|Nothing| if the key is not found. The last three functions have an \verb|Ord k| constraint, meaning that they require the key type to have a suitable total order.

\begin{figure}[h]
\begin{verbatim}
type Map k v
empty :: Map k v
insert :: Ord k => k -> v -> Map k v -> Map k v
delete :: Ord k => k -> Map k v -> Map k v
lookup :: Ord k => k -> Map k v -> Maybe v
\end{verbatim}
\caption{An excerpt from the \texttt{Map} module.}
\label{fig:maps-types}
\end{figure}

QuickSpec supports testing polymorphic functions, but its support for typeclass polymorphism is immature. To test the map API, we will therefore specialise the map functions to a specific type of keys and values. We start by defining two types \verb|Key| and \verb|Val|, which are thin wrappers around integers:
\begin{verbatim}
newtype Key = Key Int deriving (Eq, Ord, Arbitrary)
newtype Val = Val Int deriving (Eq, Ord, Arbitrary)
\end{verbatim}

Then we can define a signature, which is shown in Figure~\ref{fig:maps-sig}. We note three things about this signature:
\begin{itemize}
    \item When a function involves a user-defined type, we must declare that types in the signature. Here, we use the command \verb|monoTypeWithVars| to declare the types \verb|Key|, \verb|Val| and \verb|Map Key Val|.
    The string argument (\verb|"k"|, \verb|"v"|, \verb|"m"|) specifies what names should be used for variables of that type when printing conjectures.
    \item Because \verb|lookup| returns a \verb|Maybe Val|, its laws may involve the \verb|Maybe| constructors \verb|Just| and \verb|Nothing|, which we therefore add to the signature. However, we are not interested in \verb|Just| and \verb|Nothing| in themselves, but only in their interactions with the map operations. We therefore declare them as \emph{background functions} using the \verb|background| combinator. Background functions can appear in conjectures but any conjecture must involve at least one non-background function.
    \item The \verb|A| in the type signatures for \verb|Nothing| and \verb|Just| is a polymorphic \emph{type variable}. Polymorphic functions can be declared in QuickSpec by using the types \verb|A|, \verb|B|, \verb|C| etc.\ in type declarations. QuickSpec will then take care of instantiating the functions at appropriate types.
\end{itemize}

\begin{figure}[h]
\begin{verbatim}
mapsSig =
  [monoTypeWithVars ["k"] (Proxy :: Proxy Key),
   monoTypeWithVars ["v"] (Proxy :: Proxy Val),
   monoTypeWithVars ["m"] (Proxy :: Proxy (Map Key Val)),
   background [
     con "Nothing" (Nothing :: Maybe A),
     con "Just" (Just :: A -> Maybe A)],
   series [sig1, sig2, sig3]
   where
     sig1 =
       [con "empty" (Map.empty :: Map Key Val),
        con "lookup" (Map.lookup :: Key -> Map Key Val -> Maybe Val)]
     sig2 =
       [con "insert" (Map.insert :: Key -> Val -> Map Key Val -> Map Key Val)]
     sig3 =
       [con "delete" (Map.delete :: Key -> Map Key Val -> Map Key Val)]
\end{verbatim}
\caption{A signature for maps.}
\label{fig:maps-sig}
\end{figure}

The reader might wonder why we go to the trouble of defining the types \verb|Key| and \verb|Val|, rather than simply using \verb|Map Int Int|. We in fact used \verb|Map Int Int| at first, but got several strange properties in which the same variable appeared both as a key and a value, such as the following:
\begin{verbatim}
lookup x (insert y y m) = lookup x (insert y x m)
insert x y (insert y x m) = insert y x (insert x y m)
insert x x (insert y y m) = insert y y (insert x x m)
\end{verbatim}
Because these properties mix up keys and values, they are not useful. By keeping \verb|Key| and \verb|Val| distinct, we avoid generating them.
In general, we have found that QuickSpec finds better properties when conceptually different types are kept distinct.

Running \verb|quickSpec mapsSig| produces the following 11 conjectures:
\begin{verbatim}
  1. lookup k empty = Nothing    
  2. lookup k (insert k v m) = Just v
  3. lookup k (insert k2 v empty) = lookup k2 (insert k v empty)
  4. insert k v (insert k v2 m) = insert k v m
  5. insert k v (insert k2 v m) = insert k2 v (insert k v m)
  6. delete k empty = empty      
  7. lookup k (delete k m) = Nothing
  8. delete k (delete k2 m) = delete k2 (delete k m)
  9. delete k (delete k m) = delete k m
 10. delete k (insert k v m) = delete k m
 11. insert k v (delete k m) = insert k v m
\end{verbatim}

These properties are quite revealing of how the map API works. For example, what does the \verb|insert| function do when the key to be inserted is already present in the map? It might conceivably throw an exception, or leave the map unchanged. But property 4 reveals that the new value overwrites the old value. Similarly, property 9 shows that deleting a key which is not present does nothing. Properties 10 and 11 show that each of \verb|delete| and \verb|insert| undoes the effect of the other. Property 2 shows that \verb|lookup| retrieves a value added by \verb|insert|. Property 8 shows that \verb|delete| commutes with itself: when deleting two keys from the map, the order in which we delete the pairs doesn't matter.

We might wonder why there is no similar commutativity property for \verb|insert|---that is, a conjecture stating that, when we add two key-value pairs to the map, the order in which we add the pairs doesn't matter. Formally, this can be expressed as the property
\begin{verbatim}
insert k v (insert k2 v2 m) = insert k2 v2 (insert k v m)
\end{verbatim}
The reason this property is not found is because it doesn't hold! If the two keys are the same, that is \verb|k = k2|, then the order of insertion \emph{does} matter, because the value which is inserted last takes precedence. The property holds only if \verb|k /= k2|.

QuickSpec can discover such conditional properties, but we must explicitly declare what conditions we are interested in. The condition we need here is inequality between keys. We declare it by adding the following line to the signature, in the background function section:
\begin{verbatim}
predicate "/=" ((/=) :: Key -> Key -> Bool)    
\end{verbatim}
Now QuickSpec will discover properties having e.g. \verb|k1 /= k2| as a precondition. It will test these properties by generating \emph{random} keys and discarding any test cases where \verb|k1 = k2|.\footnote{Sometimes, this generation strategy produces bad quality test data. When that happens, the user can provide their own test data generator by using the operator \texttt{predicateGen}, a variant of \texttt{predicate} which accepts a generator as an argument.}

Before re-running QuickSpec, we need to make two small adjustments. Firstly, QuickSpec by default tests its conjectures on 1000 randomly-generated test cases. However, conditional properties can be hard to falsify, and 1000 tests is not always enough. We increase the number of tests to 10000, by adding the declaration
\verb|withMaxTests 10000| to the signature, e.g. below the \verb|monoTypeWithVars| declarations.

Secondly, QuickSpec only discovers properties up to a given \emph{size limit}. By default, properties with at most 7 symbols (variables and functions) on each side are discovered. However, when discovering conditional properties, each precondition adds 1 to the computed size. We therefore need to tell QuickSpec to search for slightly bigger properties. To increase the size limit to 8 symbols, we add the declaration \verb|withMaxTermSize 8| to the signature.

Having done that, we re-run QuickSpec, and it produces the following 4 conditional conjectures, as well as 5 unconditional equations of size 8 which we do not reproduce here:
\begin{verbatim}
12. k /= k2 => lookup k (insert k2 v m) = lookup k m
13. k2 /= k => insert k v (insert k2 v2 m) = insert k2 v2 (insert k v m)
14. k /= k2 => lookup k (delete k2 m) = lookup k m
15. k /= k2 => insert k v (delete k2 m) = delete k2 (insert k v m)
\end{verbatim}

Properties 12 and 14 show that the result of \verb|lookup| is not affected by inserting or deleting an unrelated key. Properties 13 and 15 show that \verb|insert| commutes with both itself and \verb|delete|, when the two keys involved are different. (The fact that \verb|delete| commutes with itself was found earlier as the unconditional property 8.)

In fact, these properties form a \emph{complete specification} for the map operations. 
This is because we can use them to transform any particular sequence of calls of \verb|insert| and \verb|delete|, starting from \verb|empty|, into a \emph{normal form} where only \verb|insert| is used, each key is inserted once and the keys are inserted in ascending order. For example:
\begin{verbatim}
  insert 3 'a' (delete 4 (insert 3 'b' (insert 2 'c' empty)))
= { property 15, twice }
  insert 3 'a' (insert 3 'b' (insert 2 'c' (delete 4 empty)))
= { property 6 }
  insert 3 'a' (insert 3 'b' (insert 2 'c' empty))
= { property 4 }
  insert 3 'a' (insert 2 'c' empty)
= { property 13 }
  insert 2 'c' (insert 3 'a' empty)
\end{verbatim}
Any call to \verb|lookup| for a map of this form can be rewritten to either a \verb|Just| or \verb|Nothing|. For example, if we apply \verb|lookup 3| to the map above, we get \verb|Just 'a'| by properties 12 and 2.

Note that QuickSpec is not able to see that the specification is complete. Rather, we checked this by hand. The approach is similar to that used in \cite{monadic-quickcheck}.

\subsection{The Soundness of Music}
\label{sec:music}

In Chapter 20 of his Haskell textbook \cite{hudak}, Hudak
designs a library of musical structures.
The core of this library is the \texttt{Music} datatype below. It consists of primitive entities (notes and rests), operations to combine musical structures (parallel and sequential composition) and an operation to change the tempo of a piece of music. The full library includes operations to change the pitch or instrument of a piece of music, which we omit here.

\begin{verbatim}
data Music =
    Note Pitch Rational   -- Note p d plays note p for duration d
  | Rest Rational         -- Rest d is a rest of duration d
  | Music :+: Music       -- m1 :+: m2 plays m1 and m2 in sequence
  | Music :=: Music       -- m1 :=: m2 plays m1 and m2 simultaneously
  | Tempo Rational Music  -- Tempo x m plays m sped up by a factor of x
  | ...
\end{verbatim}

For example, the expression \verb|Note (C,4) 1| represents a middle C with a duration of one whole note.
\texttt{Note~(C,4)~1 :=: Note~(E,4)~1 :=: Note~(G,4)~1}
represents a C major chord (C, E and G).
Finally, 
\texttt{(Note~(C,4)~1 :=: Note~(E,4)~1 :=: Note~(G,4)~1) :+:
Rest 1 :+: Note~(D,4)~1} represents a C major chord, followed by one whole note of silence, followed by a D. Hudak shows how to construct increasingly intricate pieces of music using these combinators.

In Chapter 21 of his book, Hudak defines and proves the algebraic properties of his musical structures. Can QuickSpec discover these properties automatically?
It can---but getting it to do so required a good deal of thought. In this section, we report on our experiences using QuickSpec on the \verb|Music| datatype. We start by showing what QuickSpec discovered, and then describe the problems we encountered.

\begin{figure}[htbp]
\begin{center}
\begin{verbatim}
 Note :: (PitchClass, Int) -> NonNeg -> Music
 Rest :: NonNeg -> Music
(:+:) :: Music -> Music -> Music
  1. m :+: Rest 0 = m              
  2. Rest 0 :+: m = m              
  3. (m :+: n) :+: o = m :+: (n :+: o)
  4. Rest x :+: Rest y = Rest (x + y)
  5. Note p 0 :+: Note q 0 = Note q 0 :+: Note p 0
  6. Note p 0 :+: Note p 0 = Note p 0

(:=:) :: Music -> Music -> Music
  7. m :=: n = n :=: m             
  8. m :=: m = m
  9. m :=: Rest 0 = m              
 10. m :=: (m :+: n) = m :+: n      
 11. (m :=: n) :=: o = m :=: (n :=: o)
 12. m :=: Note p 0 = Note p 0 :+: m 
 13. Rest x :=: Rest y = Rest (max x y)
 14. Rest x :=: Note p x = Note p x  
 15. (m :+: n) :=: (m :+: o) = m :+: (n :=: o)
 16. m :=: (n :+: (m :+: o)) = (m :=: (n :+: m)) :+: o
 17. m :=: ((m :=: n) :+: o) = (m :=: n) :+: o
 18. Rest x :=: (m :+: Rest x) = m :+: Rest x

Tempo :: Pos -> Music -> Music
 19. Tempo 1 m = m                 
 20. Tempo x (Rest 0) = Rest 0     
 21. Tempo x (Tempo y m) = Tempo y (Tempo x m)
 22. Tempo (x * y) m = Tempo x (Tempo y m)
 23. Tempo x (Note p 0) = Note p 0   
 24. Tempo x (Rest y) = Rest (y / x) 
 25. Tempo x (Note p x) = Note p 1   
 26. Tempo x (Note p y) = Note p (y / x)
 27. Tempo x m :+: Tempo x n = Tempo x (m :+: n)
 28. Tempo x m :=: Tempo x n = Tempo x (m :=: n)
 29. Tempo (x / y) (Note p 1) = Tempo x (Note p y)
 30. Rest x :+: Tempo y (Rest z) = Tempo y (Rest z) :+: Rest x
 31. Tempo x (Rest (y + x)) = Rest 1 :+: Tempo x (Rest y)
 32. Tempo x (Rest (max y x)) = Rest 1 :=: Tempo x (Rest y)
 33. Tempo x (m :+: Rest x) = Tempo x m :+: Rest 1
 34. Tempo x (Rest x :+: m) = Rest 1 :+: Tempo x m
 35. Tempo x (m :=: Rest x) = Rest 1 :=: Tempo x m
\end{verbatim}
\caption{Complete QuickSpec output on the music example.}
\label{fig:music}
\end{center}
\end{figure}

The final output of QuickSpec is shown in Figure~\ref{fig:music}. It includes all of Hudak's axioms (they are properties 1, 2, 3, 7, 9, 11, 19, 20, 21, 22, 27 and 28), as well as other interesting and not-so-interesting properties. By studying QuickSpec's output, we can discover that:

\begin{itemize}
\item \verb|Rest 0| is an identity for both \verb|:+:| and \verb|:=:| (properties 1, 2 and 9), and is unaffected by \verb|Tempo| (property 20). Thus a zero-length rest represents an empty piece of music.

\item \verb|:+:| and \verb|:=:| are associative (properties 3 and 11), and \verb|:=:| is also commutative and idempotent (properties 7 and 8).
\verb|:+:| distributes over \verb|:=:| on the left (property 15), but not on the right. (This is because, in the expression \verb|(m :=: n) :+: o|, the piece \verb|o| starts playing only once \emph{both} \verb|m| and \verb|n| are finished.)

\item Consecutive rests can be combined (property 4), as can consecutive zero-length notes of the same pitch (property 6). Parallel rests can be combined by taking the one with the longest duration (property 13). A rest played in parallel with a note of the same length can be ignored (property 14).

Consecutive notes of the same pitch can \emph{not} be combined if they have non-zero length. (Musically, repeating a note sounds different than playing it continuously.)

\item Multiplying a piece's tempo by 1 has no effect (property 19). Multiplying the tempo by $xy$ is the same as multiplying it by $x$ and then by $y$ (property 22). Multiplying the tempo of a note or rest by $x$ reduces its duration by a factor of $x$ (properties 24 and 26). Finally, \verb|Tempo| distributes over \verb|:+:| and \verb|:=:| (properties 27 and 28).

These properties constitute a complete specification of \verb|Tempo|. This is because we can use them to \emph{eliminate} any use of \verb|Tempo| from a piece of music: first use properties 27 and 28 to push \verb|Tempo| inwards until it reaches a note or a rest, then use properties 24 and 26 to remove it.
\end{itemize}

These properties reveal a good deal about how the \verb|Music| datatype works, and include all of Hudak's handmade axioms. However, setting QuickSpec up to find them took some care. We now describe the steps we went through to find these properties.

\sloppy
\paragraph{Random Generator}
To use QuickSpec on a custom datatype such as \verb|Music|, we must define an \verb|Arbitrary| instance, which describes how to generate random values of that type. Our generator was defined in a standard way, and we do not discuss it here; see \cite{quickcheck} for more information.

\fussy
\paragraph{Observational Equivalence}
Hudak defines two musical objects to be equal if they sound the same when performed.
Formally, the music library provides an operator
\begin{verbatim}
perform :: Context -> Music -> Performance
\end{verbatim}
which transforms a piece of music into the specific sequence of notes that should be played, together with their timings. (Here, \verb|Context| encodes various settings such as the speed of the performance.) Hudak then defines two pieces $m_1$ and $m_2$ to be equivalent if $\forall c.\, \texttt{perform}\ c\ m_1 = \texttt{perform}\ c\ m_2$.

This is an \emph{observational equivalence}, where two objects are considered equal if they produce the same result in all contexts. By default, QuickSpec uses the built-in Haskell operator \verb|==| for equality testing, which for \verb|Music| means structural equality. However, we can also configure QuickSpec to use observational equality. To do so, we must first declare the type in the signature using the following syntax:
\begin{verbatim}
monoTypeObserveWithVars ["m"] (Proxy :: Proxy Music).
\end{verbatim}
We then define our observational equality function by making an instance of the \verb|Observe| type class:
\begin{verbatim}
instance Observe Context Performance Music where
  observe context m = perform context m
\end{verbatim}
To test $m_1 = m_2$, QuickSpec will pick random \verb|Context|s $c$, and test that $\texttt{perform}\ c\ m_1 = \texttt{perform}\ c\ m_2$ holds for all $c$. In general, the first parameter to \verb|Observe| is a \emph{context} in which we can evaluate the object, and the second parameter is the \emph{result} we get upon evaluating it.

\paragraph{Preconditions}
At this point, we ran QuickSpec, and were greeted with a curious \emph{lack} of properties. There were \emph{no properties at all} about \verb|:+:|, not even associativity. Testing with QuickCheck revealed why not:
\begin{verbatim}
> quickCheck (\m1 m2 m3 -> m1 :+: (m2 :+: m3) =~= (m1 :+: m2) :+: m3)
*** Failed! Exception: 'Ratio has zero denominator' (after 4 tests)
\end{verbatim}
It appears that we encountered a division by zero. The reason is that in \verb|Tempo|, the speed argument must be a \emph{positive} number---it is not allowed to multiply the speed by zero (or indeed by a negative number). Thus, although \verb|Tempo| is declared as accepting any \verb|Rational| argument, it has a hidden precondition.
What's more, \verb|Note| and \verb|Rest| also have a precondition---their \verb|Rational| argument must be non-negative.

We would like to tell QuickSpec to only apply \verb|Tempo| to positive arguments, and \verb|Note| and \verb|Rest| to non-negative arguments. Unfortunately, this is not currently possible. Instead, we modified the music API to expose these preconditions in the types. We first defined types of positive and non-negative rationals:
\begin{verbatim}
newtype Pos = Pos Rational deriving (Eq, Ord, Show, Num, Fractional)
newtype NonNeg = NonNeg Rational deriving (Eq, Ord, Show, Num, Fractional)
\end{verbatim}
Then we defined versions of \verb|Tempo|, \verb|Note| and \verb|Rest| which take a \verb|Pos| or \verb|NonNeg| argument:
\begin{verbatim}
tempo :: Pos -> Music -> Music
note :: Pitch -> NonNeg -> Music
rest :: NonNeg -> Music

tempo (Pos x) m = Tempo x m
note p (NonNeg x) = Note p x
rest (NonNeg x) = Rest x
\end{verbatim}
We used these in our signature instead of the original constructors.

\paragraph{Observational Equivalence Again}
When we re-ran QuickSpec, it indeed found that \verb|:+:| is associative, along with many other properties. But something was still strange. Among the properties discovered were these two:
\begin{verbatim}
Rest x = Rest y
m :+: Rest x = m
\end{verbatim}
The first states that all rests are equivalent, which must surely be wrong. The second states that adding a rest to the end of a piece of music has no effect. This sounds reasonable, as the extra rest cannot be heard. But it is wrong, because \verb|(m :=: Rest x) :=: n| is not equivalent to \verb|m :=: n|: in the first, there is a gap of \verb|x| whole notes between \verb|m| and \verb|n|, while in the second, there is no gap.

The problem was that our \emph{observational equivalence} was defined incorrectly. It is \emph{not} true that two pieces are equivalent if they sound the same, because the pieces may end with different lengths of silence. To detect silence at the end of a piece, we modified our observational equivalence test so that it \emph{adds an extra note at the end of the piece} before performing it. The resulting \verb|Observe| instance looks as follows:
%
\begin{verbatim}
instance Observe Context Performance Music where
  observe context m = perform context (m :+: Note (C, 4) 1)
\end{verbatim}

\paragraph{Success}
Finally, we re-ran QuickSpec and it worked correctly, producing the properties shown in Figure~\ref{fig:music}.
Figure~\ref{fig:music-sig} shows the final signature we used. We note that:
\begin{itemize}
\item We declared the \verb|Music| type as using observational equality.
\item We included arithmetic operations for \verb|Pos| and \verb|NonNeg|. Rather than declaring these operations by hand, we used \verb|arith|, a pre-made signature provided by QuickSpec which declares \verb|0|, \verb|1| and \verb|+|. 
Since 0 is not positive, we excluded it for \verb|Pos| by using the \verb|without| combinator.
(Another useful pre-made signature is \texttt{lists}, which declares some common list functions.)
\item We included multiplication and division for tempos, and maximum for durations, as we guessed that they might satisfy interesting properties.
\item To encode the fact that every positive number is non-negative, we added an injection function from \verb|Pos| to \verb|NonNeg|. We used \verb|""| for the name of the function, which makes it invisible in properties.
\end{itemize}

\begin{figure}[h]
\begin{verbatim}
musicSig = [
  monoType (Proxy :: Proxy NonNeg),
  monoType (Proxy :: Proxy Pos),
  monoTypeWithVars ["p", "q", "r"] (Proxy :: Proxy Pitch),
  monoTypeObserveWithVars ["m", "n", "o"] (Proxy :: Proxy Music),
  background [
    arith (Proxy :: Proxy NonNeg),
    arith (Proxy :: Proxy Pos) `without` ["0"],
    con "*" ((*) :: Pos -> Pos -> Pos),
    con "/" ((/) :: Pos -> Pos -> Pos),
    con "max" (max :: NonNeg -> NonNeg -> NonNeg),
    con "" (\(Pos x) -> NonNeg x)]
  series [sig1, sig2, sig3]]
  where
    sig1 =
      [con "Note" note,
       con "Rest" rest,
       con ":+:" (:+:)]
    sig2 = [con ":=:" (:=:)]
    sig3 = [con "Tempo" tempo]
\end{verbatim}
\caption{A signature for the music library.}
\label{fig:music-sig}
\end{figure}

\paragraph{Conclusion}
Getting good results from QuickSpec for the music library required some care. We had to define a notion of observational equality, and found that the obvious definition was wrong. We also had to tighten the API to avoid creating pieces with negative duration or infinite tempo. Once we did that, QuickSpec was able to produce high-quality conjectures for the music library, finding all of Hudak's axioms.

\section{Combining Theorem Provers and Theory Exploration}
QuickSpec was originally designed as a stand-alone tool for automatically discovering equational conjectures about functional programs written in Haskell, as shown above. In itself it only relies on testing, but it has been connected to several automated theorem provers and successfully used as a lemma discovery tool. 
 The HipSpec system was designed for automating proofs by induction \cite{hipspec} and used QuickSpec to generate potential lemmas for the main conjecture given by the user. HipSpec then tried to prove these lemmas as well, and any lemma proved could be used in subsequent proof attempts of other lemmas and the main conjecture. This allowed improvements in automation beyond the previous state of the art. HipSpec was a light-weight prover: it simply applied (structural) induction to conjectures and sent off the resulting subgoals to an external automated theorem prover. 
In essence, HipSpec uses QuickSpec to construct a richer background theory for the inductive proofs, making them feasible to prove automatically. This is also useful in interactive proof assistants when beginning a new formalisation of some stucture about which few theorems have been proved: QuickSpec has been combined with Isabelle/HOL in the Hipster system \cite{hipster,hipster2}, and similarly also with HOL4 \cite{hopster}. Here, conjectures proposed by QuickSpec can be proved automatically by custom tactics, or if unsuccessful, left open for the user to prove interactively.

\subsection{Facilitating Interoperability: TIP and TIP-Tools}
We have developed a suite of tools to make it easier to combine QuickSpec with various theorem provers and other systems (\url{http://tip-org.github.io}). These are based around a language format called TIP \cite{tip}, which is a very slight extension of SMTLIB \cite{smtlib}, and associated tools for converting problems and conjectures to and from TIP and other common formats for theorem provers \cite{tip-tools}, using the command line tool \texttt{tip}. We also provide a TIP front-end for calling QuickSpec, using the command line tool \texttt{tip-spec}. This makes it is possible to call QuickSpec on files given in TIP format, which is used in for example the Hipster system mentioned above. Hipster first translates from Isabelle to TIP, then passes the TIP file to QuickSpec (using the \texttt{tip-spec} tool) and finally translates the resulting conjectures back to Isabelle/HOL format. Note that when calling QuickSpec via the \texttt{tip-spec} front end, the generator functions, used to create random values for testing, are automatically derived, based on the datatypes. This means that the user has a little less flexibility than when using QuickSpec directly through Haskell. 
In the next section, we demonstrate how a new theorem prover can easily be combined with QuickSpec using the TIP-tools.

\subsection{An Experiment: Combining QuickSpec and Vampire}
\begin{figure}
    \centering
    \includegraphics[scale=0.28]{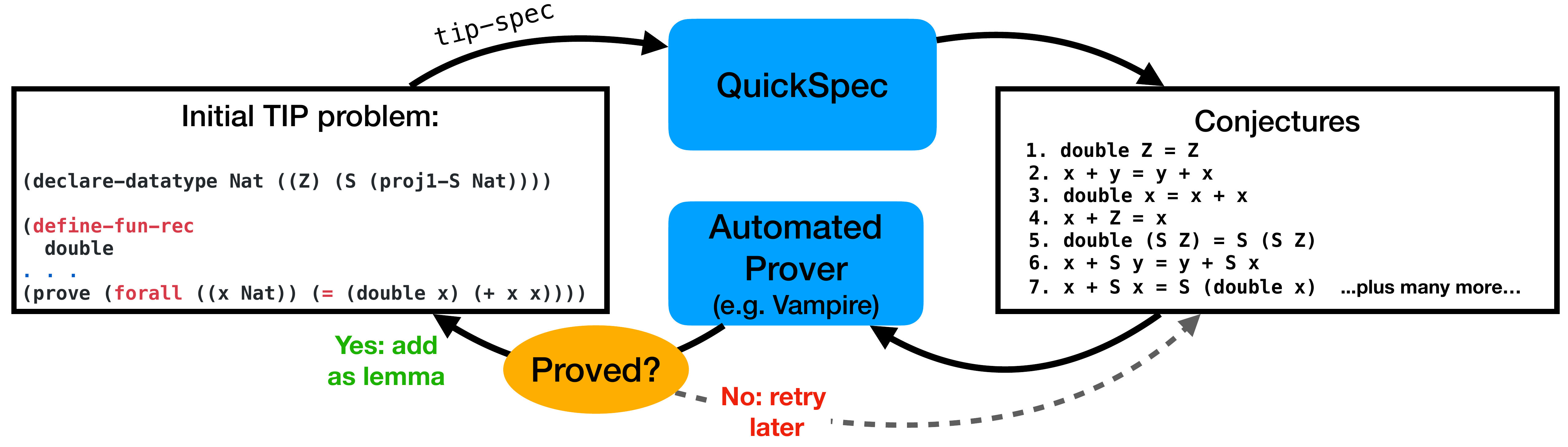}
    \caption{QuickSpec explores the functions and types in the original problem and produce a set of conjectures. Any lemmas proved may be used when finally trying the original conjecture.}
    \label{fig:explore}
\end{figure}

How easy is it to combine theory exploration with an off-the-shelf theorem prover? 
The automated first-order prover Vampire has recently been extended to support proofs by induction \cite{vampind}, although with a focus on conjectures over the integers and with many repeated variables. Vampire does not support lemma discovery as such, but has techniques for generalising conjectures which can be useful if conjectures contain repeated variables or subterms. 
We wanted to test if we could combine QuickSpec and Vampire, in order to carry out a small experiment to demonstrate the importance of auxiliary lemmas for many inductive proofs, on a set of benchmarks somewhat different from those that Vampire's induction has previously been evaluated on. We use a very similar architecture for combining QuickSpec and a prover as used previously in e.g. HipSpec, shown in Figure~\ref{fig:explore}.

For the experiment, we used Vampire 4.5.1 on a set of benchmarks from the TIP repository \cite{tip}, specifically compiled for testing inductive provers on proofs about recursive datatypes, and often requiring auxiliary lemmas. These benchmarks were originally presented in \cite{productiveuse}. The benchmark set contains 50 theorems and is available online in TIP format\footnote{\url{https://github.com/tip-org/benchmarks/tree/master/benchmarks/prod}}.
As the TIP benchmarks are written in a format allowing pattern matching in recursive function definitions, we first translate them to the SMT-LIB format accepted by Vampire, with axiomatised function definitions\footnote{Using the command: \texttt{tip \textlangle{}filename\textrangle{} --axiomatize-funcdefs2 --smtlib}}. 

Without theory exploration, Vampire\footnote{Using the best-performing version from \cite{vampind} with structural induction and generalisation turned on:\\ \texttt{vampire_rel_4.5.1 -ind struct -indgen on}} proves 7 out of 50 conjectures, as shown in Table \ref{tab:vamp}. Next, we ran QuickSpec as a pre-processing step. For each TIP problem file, QuickSpec explored the functions occurring in the problem to produce extra conjectures. Vampire was then given the conjectures to prove, and any it managed to prove were allowed to be used as lemmas in the proof of the main theorem (see Figure~\ref{fig:explore}). This was all implemented in just a 90-line shell script.
Vampire now proved an additional 18 benchmarks, shown in Table \ref{tab:QSvamp}. Note that for three problems (properties 33--35) QuickSpec timed out during testing, as the functions (e.g. exponentiation recursively defined on natural numbers) produced very large test terms unless custom settings were used.
In total Vampire proved 25 benchmarks. As a comparison, the same benchmarks were used to evaluate several inductive theorem provers in \cite{hipspec}; here HipSpec proved 44 problems (excluding the three needing custom settings), CLAM \cite{productiveuse} was reported to have proved 41 fully automatically (and the rest interactively), and Zeno \cite{zeno} proved 21. While Vampire's induction is not tailored specifically to these kinds of problems, its performance is still reasonable compared to dedicated inductive provers, when combined with QuickSpec for lemma discovery.  
\begin{table}[htbp]
\begin{tabular}{cl|cl}
  \multicolumn{2}{l}{Prop.} &  \multicolumn{2}{l}{Prop.}  \\
 \hline
 2 & \textit{length (xs ++ ys) = length (ys ++ xs)} & 43 & \textit{elem x ys $\Rightarrow$ elem x (union zs ys) }\\
 3 & \textit{length (xs ++ ys) = (length ys) + (length xs)} & 45 & \textit{elem x (insert x ys) }\\
 15 & \textit{x + (Suc x) =  Suc (x + x)}  & 46  & \textit{x = y $\Rightarrow$ elem x (insert y zs)}   \\
 37 & \textit{(elem x zs) $\Rightarrow$ elem x (ys ++ zs)} &  & 
\end{tabular}
\caption{Properties proved by Vampire (without extra lemmas from QuickSpec).}
\label{tab:vamp}
\end{table}

\begin{table}[htbp]
\begin{tabular}{cl|cl}
  \multicolumn{2}{l}{Prop.} &  \multicolumn{2}{l}{Prop.}  \\
 \hline
 1 & \textit{double x = x + x } &                       18  & \textit{rev ((rev xs)++ ys) = (rev ys) ++ xs}\\
 4 &  \textit{length (xs ++ xs) = double (length xs)} & 19 & \textit{(rev (rev xs)) ++ ys = rev (rev (xs ++ ys)) } \\
 5 &  \textit{length (rev xs) = length xs } &           20 & \textit{even (length (xs ++ xs)) } \\
 6 & \makecell[vl]{\textit{length (rev (xs ++ ys)) =} \\ \quad \textit{(length xs) + (length ys) }} &   22  &\makecell[vl]{ \textit{even (length (xs ++ ys)) =} \\ \quad \textit{even (length (ys ++ xs))}} \\
 
 10 & \textit{rev (rev xs) = xs} &                   23 &\makecell[vl]{\textit{half (length (xs ++ ys)) =} \\ \quad \textit{half (length (ys ++ xs)) }}\\
 
 11 & \textit{rev (rev xs ++ rev ys) = ys ++ xs } &   24 & \textit{even (x + y) = even (y + x)} \\
 13 &  \textit{half (x + x) = x } &                   25 & \makecell[vl]{\textit{even (length (xs ++ ys)) =} \\ \quad \textit{even ((length ys) + (length xs)) } }\\
 16 &  \textit{even (x + x)} &                  26 & \textit{half (x + y) = half (y + x)} \\
 17 & \makecell[vl]{\textit{rev (rev (xs ++ ys)) =} \\ \quad \textit{(rev (rev xs)) ++ (rev (rev ys))} }&   30 & \textit{rev ((rev xs) ++ []) = xs } \\
 
\end{tabular}
\caption{Additional properties proved by Vampire with help of lemmas from QuickSpec.}
\label{tab:QSvamp}
\end{table}

\section{Related Work in Theory Exploration and Automated Conjecturing}
Conjecture generation systems tend to roughly fall into three categories: heuristic rule-based systems, term generation-and-testing (to which QuickSpec belongs) and 
neural network-based systems.

\paragraph{Rule-based and Heuristic}
The AM system is probably the first system designed to discover mathematical concepts \cite{AM}. It relied on several hundred heuristic rules to generate mathematical concepts and conjectures, with each rule contributing a pre-defined ``interestingness score'' to the result. 
 The HR system had a similar goal of automating concept formation and conjecturing  \cite{HR}. It took a set of axioms as inputs, and combined model finding with heuristic production rules to invent new conjectures of interest. The MATHsAiD system was specifically designed as a theory exploration system for mathematicians \cite{mathsaid}. The discovery process was guided by a forward reasoning process, which instantiated templates called \emph{theorem shells}, following heuristic plans constructed to reflect human mathematical reasoning.

In comparison, QuickSpec is a more light-weight system concerned only with conjecturing, not any other concept formation tasks. It does not employ explicit heuristic rules: it simply generates terms about the given concepts, and evaluates them on random values to learn which ones appear to be equal. It also uses a very simple notion of interestingness: a term is interesting to explore if it is non-reducible with respect to what is known so far. This makes it more flexible, as it can be used to explore conjectures about any datatype for which it can generate examples.

\paragraph{Term Generation and Testing}
 Graffiti was a conjecture generation system specifically for graph theory  \cite{graffiti}. It maintained a library of example graphs, and attempted to generate conjectures of certain shapes, then checking them for consistency with its library examples. Unlike QuickSpec, where new examples can be generated on demand, Graffiti's set of examples was fixed, leading to production of relatively many non-theorems.
 The systems IsaCoSy \cite{isacosy} and IsaScheme \cite{isascheme} are conceptually similar to QuickSpec, relying on term generation with different restrictions to avoid trivial and redundant conjectures, combined with random testing. A key difference is however that QuickSpec evaluates terms (and records the results), not whole equational statements, which greatly improves efficiency. Another closely related system from the functional programming community is Speculate \cite{speculate}, which like QuickSpec, was designed to discover properties about Haskell programs.  
 
\paragraph{Neural Networks}
Recently, techniques using large neural network architectures from natural language processing have been applied to the problem of discovering new conjectures. Urban et al. train the transformer model GPT-2 on the Mizar Mathematical Library \cite{urban2020}. With the right parameters, they get GPT-2 to generate novel and well-typed conjectures in Mizar format. However, the output may include duplicates from the library as well as non-theorems. 
Rabe et al. present a system based on self-supervised language models for mathematical reasoning tasks, including generating a missing precondition for a given conditional statement, generating one side of an equation given the other side, as well as ``free-form'' conjecturing \cite{rabe2021}. 
Between 13--30\% of generated statements were both provable and new, with the remainder being for instance exact copies or alpha-renamings of statements from the training set, or simply false. 

The neural network approach is of course radically different to QuickSpec. QuickSpec was designed to produce relatively small sets of interesting equations at a time, intended to be read by a human as a specification for a functional program. It rarely generates any false or duplicate conjectures, thanks to the integrated testing in the equation formation. Furthermore, QuickSpec is not dependent on training data about the theory at hand being available, but does on the other hand require generators to produce test data, with the caveat that the test data needs to be computable, or that a computable proxy can be used. 

\section{Conclusion and Further Work} 
The idea behind QuickSpec is simple: generate interesting terms and evaluate them on random test data to see which ones appear to be equal. Despite this, it is fast and efficient when given relatively small theories to explore. In this sense, QuickSpec is data-driven: it uses generator functions from the automated testing framework QuickCheck to generate and evaluate as many ground examples as it needs. This is also one of its limitations: trying to evaluate functions of high complexity (e.g. exponentiation in Peano arithmetic) can drastically slow down testing if a large test case is generated. Similarly, co-recursive functions can lead to non-terminating values \cite{cohipster}. One solution is to extend QuickSpec with observational equivalence checking, similarly as shown in the Music example in section \ref{sec:music}. 
The other main weakness of QuickSpec is scalability. As it generates all non-redundant terms, the search space eventually becomes intractable when faced with large theories (say, a library with 30+ functions) or when asked to explore very large terms (sizes over about 9). Here, a combination with machine learning could be fruitful. A human user can immediately judge which of QuickSpec's equations are ``nice and sensible'' or, on occasion, ``weird and ugly'' and thus not very interesting. Certain shapes of conjectures are often both more useful and aesthetically pleasing, and QuickSpec could be steered towards that part of the search space first. A recent extension to QuickSpec experiments with doing this, using user-provided templates when exploring larger theories and terms \cite{RoughSpec}. A natural next step is to instead attempt to learn templates from a library of lemmas, as suggested in \cite{proofpatternrec}, thus creating a hybrid system where machine learning steers the term generation towards interesting parts of the search space. Learned templates may also facilitate search of non-equational conjectures, while still managing search space size.

\bibliographystyle{eptcs}
\bibliography{paper}
\end{document}